# Design of energy absorbing metamaterials using stochastic soft-wall billiards

V.N. Pilipchuk
Wayne State University, Detroit, USA

**Abstract**

Physical principles for designing artificial materials with energy absorbing/harvesting and wave guiding properties are discussed in the present work. The idea is to insert light particles in a lattice of elastically coupled potential wells representing soft-wall versions of the so-called stochastic billiards. A planar case of a single potential well attached to the base with a linearly elastic spring and including one or few small particles was considered earlier (https://doi.org/10.1007/s11071-019-05109-z). Here, we analyze the evolution of waves in a one-dimensional lattice of 3D potential wells with light particles inside. By assigning the initial conditions corresponding to propagating waves we found that the waves can be quickly destroyed by increasing the mass of particles while certain shapes of the potential containers provide a quasi-one-directional energy flow from the chain of containers into the chaotically moving particles by increasing the 'temperature' of the lattice.

## Introduction

Physical principles for designing energy absorbing/harvesting metamaterials[1] are discussed in the present work. We consider a lattice of elastically coupled potential wells representing 3D versions of the so-called soft wall stochastic billiards with light particles inside. A planar case of a single container attached to the base with a linearly elastic spring and including one or few small particles was introduced earlier [1] in the Lagrangian form (Fig. 1a)

$$L = \frac{1}{2}\dot{X}^2 - \frac{1}{2}X^2 + \sum_{j=1}^{k}\left\{\frac{\mu}{2}\left[\left(\dot{X} + \dot{x}_j\right)^2 + \dot{y}_j^2\right] - V(x_j, y_j)\right\}$$

$$V(x, y) = \frac{\gamma}{2n}\left[\left(\frac{x}{\alpha + \beta(y^2 - 1)}\right)^{2n} + y^{2n}\right]; \quad \alpha = 1/2, \quad \gamma = 1.0 \tag{1}$$

where $\beta$ and $n$ are parameters controlling the container's shape and the effective wall's stiffness, respectively.

The potential well $V(x,y)$ was introduced in such a way that, in the rigid-body limit, $n \to \infty$, it resembles one of the two most common types of billiards, with either dispersing [2]- or stadium-type [3] boundaries when the main geometrical parameter of the well $\beta$ changes its sign. The corresponding contour lines/boundaries are represented by the upper part of Fig. 1b for three different values of the parameter $\beta$. Furthermore, it was shown that the light particles inside the potential well can absorb the energy from the oscillating container $M$ for quite a long time by delaying the Poincaré recurrence beyond physically reasonable temporal intervals of the system. We emphasize that no phenomenological dissipation nor specific initial conditions were imposed. Nonetheless, the energy flow from the container (donor) to the inner particle (acceptor) appeared to have statistically irreversible trends whenever the dynamics of particle inside the potential well are chaotic (Fig.1b). Conditions of chaotic motions were associated with instabilities of the selected nonlinear normal mode according to definition [4]. A possible analogue of this phenomena may associate with the so-called stochastic acceleration of particles as originally explained by E. Fermi [5], although the physical nature of forces considered in cosmology is quite different.

---

[1] Generally, an artificial material with a specific property, which does not exist in natural materials.

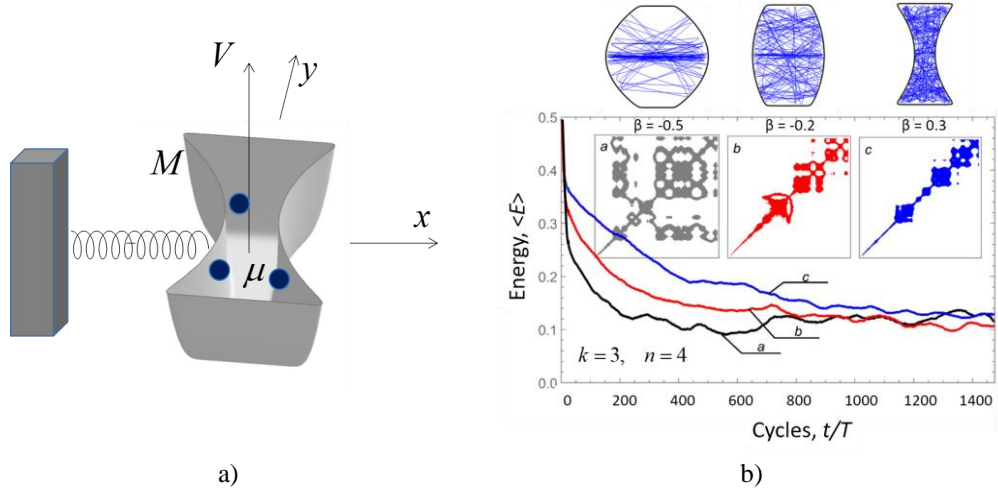

a)                                          b)

**Figure 1.** Numerical evidence of energy absorbing properties of soft-wall stochastic billiards: **a)** a massive potential container *M* oscillating with small particles interacting with potential walls of different 'stiffness' and shapes; the container mass and spring stiffness are scaled to be unity, and $\mu = m/k$, where *k* and *m* are the number of particles and their total mass, respectively; **b)** running average of the total energy of oscillating container with the corresponding recurrence diagrams, as defined in [6], for *different container shapes* shown at the top with sample chaotic particle' trajectories, where *T* is the eigen period of oscillation of the container.

## 1. A chain model of 3D soft-wall stochastic billiards

In the present work, we consider the effects of wave propagation in one-dimensional lattices of the 3D version of potential wells introduced at the previous stage of investigation. By assigning the initial conditions corresponding to propagating waves we found that the waves can quickly be destroyed by increasing the mass of particles while varying shapes of the potential containers revealed the effect of quasi-one-directional energy flow from the waves to the particles by increasing the 'temperature' of the lattice. After appropriate scaling of parameters and variables, the model under consideration (Fig.2a) is described with the Lagrangian,

$$L = \frac{1}{2}\sum_{j=1}^{S}\left\{\left(\frac{dX_j}{dt}\right)^2 + \mu\left(\left(\frac{dx_j}{dt} + \frac{dX_j}{dt}\right)^2 + \left(\frac{dy_j}{dt}\right)^2 + \left(\frac{dz_j}{dt}\right)^2\right)\right\} - \sum_{j=1}^{S}\left(\frac{1}{2}KX_j^2 + \frac{1}{2}(X_j - X_{j-1})^2 + V(x_j, y_j, z_j)\right) \quad (2)$$

where $X_j$ is the displacement of $j^{th}$ container, whose dynamics are constrained to be one-dimensional, $\{x_j, y_j, z_j\}$ are relative displacements of a particle $\mu$ measured inside containers, the stiffness of coupling between containers is assumed to be unity, whereas the stiffness of base springs is *K*, and all the potential wells of the chain are described with the same function as

$$V(x,y,z) = \frac{\gamma}{2n}\left(\left(\frac{x}{\beta(y^2+z^2-1)+1}\right)^{2n} + \left(\frac{y}{\beta(x^2+z^2-1)+1}\right)^{2n} + \left(\frac{z}{\beta(x^2+y^2-1)+1}\right)^{2n}\right)$$

where the meaning of parameters are similar to the 2D case (1).

V.N. Pilipchuk

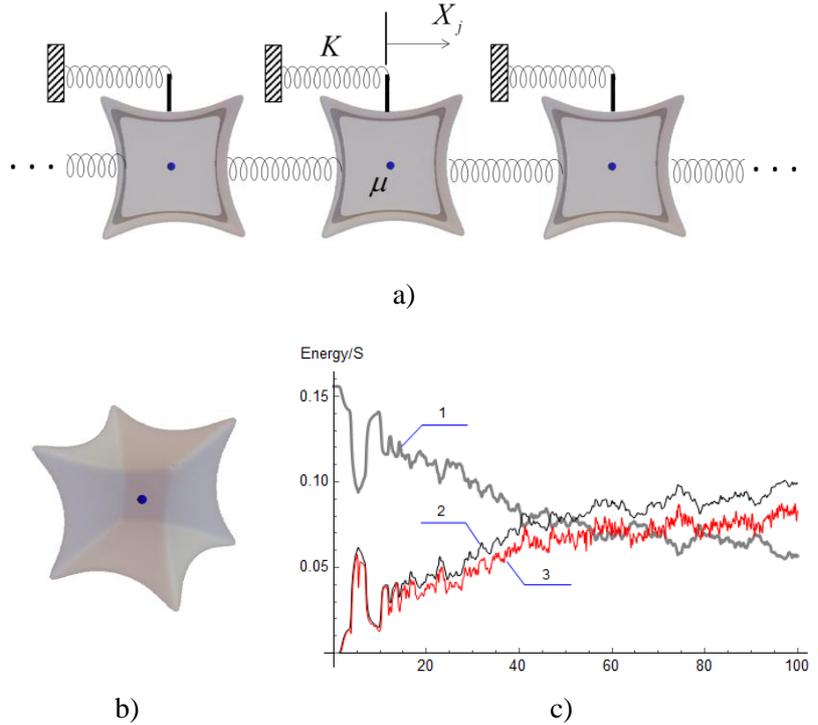

**Figure 2. a)** Chain model of soft-wall billiards, **b)** 3D image of the dispersing billiard given by the level surface $V(x, y, z) = 1$ at $\beta = 0.1$, $\lambda = 1$ and $n = 6$, **c)** time history of the total energy per cell under the initial conditions corresponding to the propagating sine wave: **1**- containers, **2**- inner particles, and **3**- kinetic energy the particles interpreted as a structural temperature; the corresponding evolution of wave shapes is illustrated below in Fig.3.

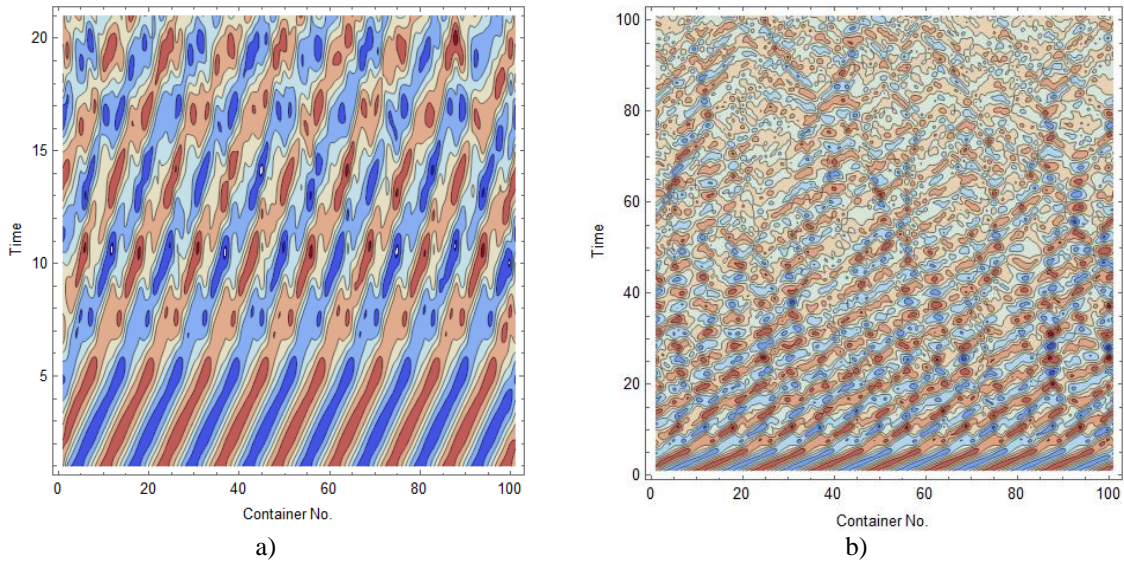

**Figure 3.** Cancellation of the propagating wave due to the quasi irreversible energy transfer from the chain of containers into its inner particles: **a)** the short-term pattern of the displacement field, and **b)** the long-term displacement field showing transition from the propagating wave to the chaotic dynamics of the chain of containers as their energy is absorbed by the inner particles (Fig.2c).

Under the condition $n \gg 1$, the container shape can be represented by the equipotential surface $V(x, y, z) = 1$, which is shown in Fig.2b for the case $\beta = 0.1$. As follows from the shape of the surface, the inner particle interacting with the soft potential walls is subjected to the repelling reaction force. The only difference with the stiff wall dispersive billiards is that the impulsive reaction forces are still continuous.

## 2. Discussion of a sample simulation result

The simulation was conducted for the chain (2) of $S = 100$ containers under the cyclical boundary conditions and the initial states corresponding to the propagating sine-wave, $X_j = -A\sin(j\lambda_k)$, $dX_j/dt = A\omega_k \cos(j\lambda_k)$ at $t=0$, where $\omega_k = \sqrt{K + 2\sin^2(\lambda_k/2)}$ and $\lambda_k = \pi k/(S+1)$. The amplitude and wave number were assigned as $A = 0.5$ and $k = 16$, respectively. We also assumed that $dx_j/dt = -dX_j/dt$ at $t = 0$ in order to keep the particles near their equilibrium states at the initial time while other initial coordinates of the particles were chosen as small random numbers from the interval $(-0.01, 0.01)$ different for different containers. The simulation was conducted under the model parameters: $K = 1.0$, $\gamma = 1.0$, $n = 6$, and $\beta = 0.1$. The result of simulation is illustrated by Fig.2c and Fig.3 to show that the wave propagates through the chain of containers with almost no disturbance until most of the inner particles reach the potential walls. Then, once the chaos of the particle motions developed, the energy flow acquires a one-directional trend from the containers to the particles (Fig.2c). Due to the chaoticity of particles and thus incoherency of this process between different containers the wave shape of container chain becomes irregular and losing the propagation property (Fig.2), although the wave collapse did not reveal a direct link to its energy loss. The latter appeared to be due to the Fermi type of acceleration of the inner particles increasing their total energy and thus structural temperature growth as most of the time the particle energy maintains its kinetic form as confirmed by comparison of curves 2 and 3 in Fig.2c.

## Conclusions

Using the chain model of 3D soft-wall billiards with dispersive boundaries, represented by massive containers, we revealed the possibility of a quasi-one-directional energy flow from the wave propagating through the chain of containers into the inner particles. Since most of the time the energy of particles preserves its kinetic form, this can be interpreted as 'heating' the structure. Also the interaction of particles with container soft walls eventually leads to a collapse of the wave propagation effect, which was found to be independent from the energy loss. Such observations can offer certain guide lines for designing artificial energy absorbing/harvesting materials on micro-, macro-, and mesoscopic levels. The model may also be adapted for analyses some natural crystal structures.